\documentstyle[12pt]{article}
\begin{document}
\begin{flushright}
Alberta-Thy-02-97\\
January, 1996\\
\end{flushright}

\begin{center}
{\Large \bf  CP   Asymmetries in  Single Isospin Channels in   B Decays  and the Role of Inelastic Final State Interactions}\\
\vskip 5mm
A.N.Kamal and C.W.Luo\\
{\em Theoretical Physics Institute and Department of Physics,\\ University of Alberta,
Edmonton, Alberta T6G 2J1, Canada.}\\
\end{center}

\vskip 5mm
\begin{abstract}
The role of inelastic final state interactions in CP asymmetries is investigated in single isospin two-body hadronic decays of the B meson. We demonstrate that in a  channel where the CP asymmetry vanishes in absence of inelastic  final state interactions, a coupling to a second channel where the asymmetry is nonvanishing can result in an asymmetry in the first channel of the same order as in the second.  
\end{abstract}

\newpage

It is well-known that CP asymmetries occur in two-body hadronic decays of B meson involving two distinct CKM angles and two different strong phases\cite{KPS1}. These strong phases could arise from perturbative penguin loops or from hadronic final state interactions (fsi) involving two different isospin states\cite{comment}.  A simple example suffices to illustrate this point.  Suppose the  exclusive  decay amplitude has the form

\begin{equation}
A=A_1e^{i (\delta^{(1)}_w+\delta^{(1)}_s)}+ A_2e^{i (\delta^{(2)}_w+\delta^{(2)}_s)}
\end{equation}
where $\delta^{(1,2)}_w $ are the weak phases arising from CKM angles and $\delta^{(1,2)}_s $
strong phases arising either from the penguin loops or from hadronic fsi, then

\begin{equation}
 \Delta \Gamma=\Gamma-\bar{\Gamma}  \sim  \sin{(\delta^{(1)}_w-\delta^{(2)}_w) }\sin{(\delta^{(1)}_s-\delta^{(2)}_s)}.  
\end{equation}
\par
If the hadronic final state involves a single isospin,  $\Delta\Gamma $ can still be nonzero. 
 For example, if  $A_1 $ were a tree amplitude ( $\delta^{(1)}_s=0 $) with a weak phase $\delta_w^{(1)} $ and $A_2 $ a penguin driven amplitude with  $\delta^{(2)}_s\neq 0 $ and a weak phase $\delta_w^{(2)} $, there could be a CP violating asymmetry in a decay involving a single isospin final state.

\par
In this report we have studied interchannel mixing and its effect on single isospin channels for B
decays. For illustrative purposes we have chosen two  pairs of decay modes: $B^-\rightarrow D^0 D^-_s $ and $ \eta_c K^- $ and $B^-\rightarrow D^0 D^- $ and $ \eta_c \pi^- $ . Both pairs involve a single isospin final states; $I=1/2 $ for the first pair and $I=1 $ for the second. For the purposes of detailed discussion, let us concentrate on the first pair.  Both decays,  $B^-\rightarrow D^0 D^-_s $ and $ \eta_c K^- $,  involve only $I=1/2 $ final states and proceed at the same level in CKM angles,  the tree diagrams being proportional to $V_{cb}V_{cs}^* $.  The former is color-favored and the latter color-suppressed.  In the absence of inelastic fsi, CP asymmetry is known \cite{KPS2} to be zero for $B^-\rightarrow \eta_c K^- $ mode while it is small but nonzero for $B^-\rightarrow D^0 D_s^- $.  We show in a model calculation that CP asymmetry of the same size as in  $B^-\rightarrow D^0 D_s^- $ can be generated in $B^-\rightarrow \eta_cK^- $ through an inelastic coupling to the channel $B^-\rightarrow D^0 D^-_s $. The same is true for the pair
$B^-\rightarrow D^0 D^- $ and $ \eta_c \pi^- $. In absence of inelastic fsi, CP violating asymmetry is large ($\sim O(few\%) $ ) in $B^-\rightarrow D^0 D^- $  but zero in  $B^-\rightarrow  \eta_c \pi^- $\cite{KPS2}. Inelastic fsi generate an asymmetry in $B^-\rightarrow  \eta_c \pi^- $
of the same order as in $B^-\rightarrow D^0 D^- $.
 
\par
The effective Hamiltonian for $b\rightarrow s $ transition is given by \cite{DH, F, BJLW}

\begin{equation}
H_{eff}={G_F  \over \sqrt{2} } \sum_{q=u,c}{\left\{  V_{qb} V^*_{qs}[ C_1 O^q_1+C_2 O^q_2+
\sum_{i=3}^{10}{C_i O_i}]  \right\} } .
\end{equation}
The operators in Eq.(3) are the following;
\begin{eqnarray}
O^q_1=(\bar{s}q)_{V-A} (\bar{q}b)_{V-A},   & O^q_2=(\bar{s}_{\alpha}q_{\beta})_{V-A} (\bar{q}_{\beta} b_{\alpha})_{V-A}; \nonumber \\
O_3=(\bar{s}b)_{V-A}\sum_{q^{\prime}} (\bar{q}^{\prime}q^{\prime})_{V-A},  & 
O_4=(\bar{s}_{\alpha} b_{\beta})_{V-A}\sum_{q^{\prime}} (\bar{q}^{\prime}_{\beta} q^{\prime}_{\alpha})_{V-A},     \nonumber  \\
O_5=(\bar{s}b)_{V-A}\sum_{q^{\prime}} {(\bar{q}^{\prime}q^{\prime})_{V+A}}, & 
O_6=(\bar{s}_{\alpha} b_{\beta})_{V-A}\sum_{q^{\prime}}{(\bar{q}^{\prime}_{\beta} q^{\prime}_{\alpha})_{V+A}};        \\
O_7={3 \over 2}(\bar{s}b)_{V-A}\sum_{q^{\prime}} {(e_{q^{\prime}}\bar{q}^{\prime}q^{\prime})_{V+A}}, & 
O_8={3\over 2}(\bar{s}_{\alpha} b_{\beta})_{V-A} \sum_{q^{\prime}}{(e_{q^{\prime}}\bar{q}^{\prime}_{\beta} q^{\prime}_{\alpha})_{V+A}},
\nonumber    \\
O_9={3 \over 2}(\bar{s}b)_{V-A}\sum_{q^{\prime}} {(e_{q^{\prime}}\bar{q}^{\prime}q^{\prime})_{V-A}}, & 
O_{10}={3\over 2}(\bar{s}_{\alpha} b_{\beta})_{V-A} \sum_{q^{\prime}}{(e_{q^{\prime}}\bar{q}^{\prime}_{\beta} q^{\prime}_{\alpha})_{V-A}}.  \nonumber  
\end{eqnarray}
$O_1 $ and $O_2 $ are the Tree operators, $O_3, ...., O_6 $ are generated by QCD Penguins and $O_7,...., O_{10} $ are generated by Electroweak Penguins.  Here  $V\pm A $ represent $\gamma_{\mu} (1\pm \gamma_5)$, $\alpha$ and $\beta$ are color indices.  $\sum_{q^{\prime}}$ is a sum over the active flavors u,d,s and c quarks.

\par
In the next to leading log calculation one works with effective Wilson coefficients $C^{eff}_i $, rather than the coefficients that appear in (3). The derivation of these effective coefficients is well known \cite{DH, F, BJLW}. We simply quote their values

\begin{eqnarray}
C^{eff}_1=\bar{C}_1,~~  C^{eff}_2=\bar{C}_2, & ~~ C^{eff}_3=\bar{C}_3 - P_s / N_c,  & C^{eff}_4=\bar{C}_4 + P_s,  \nonumber   \\
C^{eff}_5=\bar{C}_5 - P_s/N_c,~~~~ &  C^{eff}_6=\bar{C}_6 +P_s, &  C^{eff}_7=\bar{C}_7 +P_e, 
 \nonumber   \\
 C^{eff}_8=\bar{C}_8, ~~~~~~~~~~~~~~ & C^{eff}_9=\bar{C}_9 +P_e, &  C^{eff}_{10}=\bar{C}_{10}
\end{eqnarray}
with
\begin{eqnarray}
& \bar{C}_1=1.1502, ~\bar{C}_2=-0.3125, ~\bar{C}_3=0.0174, ~\bar{C}_4=-0.0373, ~\bar{C}_5=0.0104,~ \bar{C}_6=-0.0459, & \nonumber   \\
& \bar{C}_7=-1.050\times 10^{-5}, ~~\bar{C}_8=3.839\times 10^{-4}, ~~\bar{C}_9=-0.0101, 
~~\bar{C}_{10}=1.959\times 10^{-3}.~~~~~~&
\end{eqnarray}
and 
\begin{eqnarray}
P_s &= &{\alpha_s (\mu) \over  8\pi} C_1 (\mu) [{10 \over 9}+\frac{2}{3}\ln{\frac{m_q^2}{\mu^2}}-G(m_q, \mu, q^2)], \\
P_e &= &{\alpha_{em} (\mu) \over  3\pi} [C_2 (\mu)+{C_1(\mu) \over N_c}] [{10 \over 9}+\frac{2}{3}\log{\frac{m_q^2}{\mu^2}}-G(m_q, \mu, q^2)]
\end{eqnarray}
where
\begin{equation}
G(m_q, \mu, q^2)= -4\int_{0}^{1}{dx x (1-x) \ln{[1-x (1-x) {q^2 \over  m^2_q}}]},
\end{equation}
$q^2 $ is the   momentum carried by the gluon or the photon in the penguin diagram and 
$m_q$ the mass of the quark q in the penguin loop.   For  $q^2> 4 m^2_q$,  $G(m_q, \mu, q^2)$ becomes complex and a strong perturbative phase is generated. 

\par
Consider now the pair  $B^-\rightarrow D^0 D^-_s $ and $\eta_cK^- $. The CP asymmetry is known\cite{KPS2}  to be zero for $B^-\rightarrow \eta_c K^- $.  Though  \cite{KPS2} does not include the Electroweak Penguin operators, this fact does not change with their inclusion.    Ref\cite{KPS2} also calculates the CP asymmetry for $B^-\rightarrow D^0D^-_s $ to be $\sim -0.12\% $ with $q^2=m_b^2/2 $. We show below that if the channels $D^0D^-_s $ and $\eta_cK^- $ are coupled inelastically, a CP asymmetry of the same order as in $D^0D^-_s $ channel is generated in the channel $\eta_cK^-$.

\par
Let us label the channels $D^0 D^-_s $ and $\eta_cK^- $ as channels 1 and 2 respectively.
Before inelastic fsi are turned on, let the decay amplitudes for these two channels be labelled $A^{(0)}_1 $ and $A^{(0)}_2$. In the usual factorization approximation, $A^{(0)}_2 $ has no absorptive part and, thus, CP asymmetry vanishes\cite{KPS2} in $B^-\rightarrow \eta_c K^- $ channel.  Let us now couple the two amplitudes through hadronic fsi using the K-matrix method\cite{Kamal, WS} thereby generating the amplitudes $A_1$ and $A_2$,
\begin{equation}
\bf{A}= (1-i \bf{k}^{\frac{1}{2}}\bf{K}\bf{k}^{\frac{1}{2}})^{-1} \bf{A}^{(0)}.
\end{equation}
Here  $\bf{A}^{(0)} $ and $\bf{A} $ are columns with entries $A^{(0)}_1 $ and $ A^{(0)}_2 $ etc., $\bf{K} $ is a real-symmetric K-matrix and $\bf\bf{k} $ the diagonal momentum matrix,
\begin{eqnarray} 
\bf{K}=\pmatrix{
a & b \cr
b & c \cr
}, 
& ~~~~~~~~ &  \bf{k}=\pmatrix{
k_1 & 0 \cr
0 & k_2 \cr
}. 
\end{eqnarray} 
   
The corresponding S-matrix for $2\times 2 $ scattering satisfying the unitarity condition is 
\begin{equation}
\bf{S}=(1+i\bf{k}^{\frac{1}{2}}\bf{K}\bf{k}^{\frac{1}{2}}) (1-i\bf{k}^{\frac{1}{2}}\bf{K}\bf{k}^{\frac{1}{2}})^{-1}.
\end{equation}
Eq.(10) can be written explicitly as
\begin{equation}
\left (\matrix{
A_1\cr
A_2\cr
}\right )=\frac{1}{\Delta}\pmatrix{
1-ick_2 & ib\sqrt{k_1k_2} \cr
ib\sqrt{k_1k_2} & 1-iak_1 \cr
}\left (\matrix{
A^{(0)}_1\cr
A^{(0)}_2\cr
}\right ).
\end{equation}
The determinant $\Delta $ is defined later.
The two-channel S-matrix is commonly parameterized in terms of three real parameters, $\delta_1 $, $\delta_2 $ and $\eta$, the eigen phases and the elasticity,  as follows,
\begin{equation}
\bf{S}=\pmatrix{
\eta e^{2i\delta_1} & i (1-\eta^2)^{\frac{1}{2}} e^{i (\delta_1+\delta_2)} \cr
i (1-\eta^2)^{\frac{1}{2}} e^{i (\delta_1+\delta_2)} & \eta e^{2i\delta_2} \cr
}.
\end{equation}

The relation between $(\delta_1, \delta_2, \eta) $ and (a,b,c) of the K-matrix can be shown to be,
\begin{eqnarray}
a &= & \frac{sin{(\delta_1+\delta_2)} +\eta sin{(\delta_1-\delta_2)}}{k_1[cos{(\delta_1+\delta_2)} +\eta cos{(\delta_1-\delta_2)}]}, \nonumber  \\
b & = &\sqrt{ \frac{1-\eta^2}{k_1k_2}} / [cos{(\delta_1+\delta_2)} +\eta cos{(\delta_1-\delta_2)}],   \\
c &= & \frac{sin{(\delta_1+\delta_2)} -\eta sin{(\delta_1-\delta_2)}}{k_2[cos{(\delta_1+\delta_2)} +\eta cos{(\delta_1-\delta_2)}]}.  \nonumber  
\end{eqnarray}
The inverse relations are
\begin{eqnarray}
\eta & = &\sqrt{1-\frac{4k_1k_2 b^2}{|\Delta|^2}  },   \nonumber \\
2\delta_1 & = & tan^{-1}\frac{k_1 a}{\Delta_-}+ tan^{-1}\frac{k_2c}{\Delta_-}+ tan^{-1}\frac{k_1 a}{\Delta_+}- tan^{-1}\frac{k_2 c}{\Delta_+},   \\
 2\delta_2 &= &  tan^{-1}\frac{k_1 a}{\Delta_-}+ tan^{-1}\frac{k_2c}{\Delta_-}- tan^{-1}\frac{k_1 a}{\Delta_+}+ tan^{-1}\frac{k_2 c}{\Delta_+}  \nonumber  
\end{eqnarray}

where 
\begin{eqnarray}
  ~& \Delta =\Delta_{-}-i(ak_1+ck_2),   \nonumber \\
~ & \Delta_{\mp}=1\mp k_1k_2(ac-b^2).
\end{eqnarray}

\par
Using (15) in (13) it is easily shown that  Eq.(13)  can be written in terms of the three parameters $\delta_1 $, $\delta_2 $ and $ \eta $ as 
\begin{equation}
\left (\matrix{
A_1\cr
A_2\cr
}\right )=\frac{1}{2}\pmatrix{
1+\eta e^{2i\delta_1} & i (1-\eta^2)^{\frac{1}{2}} e^{i (\delta_1+\delta_2)} \cr
i (1-\eta^2)^{\frac{1}{2}} e^{i (\delta_1+\delta_2)} & 1+\eta e^{2i\delta_2} \cr
}\left (\matrix{
A^{(0)}_1\cr
A^{(0)}_2\cr
}\right ).
\end{equation}
More compactly, for two channels in K-matrix unitarization,
\begin{equation}
\bf{A}=\frac{1}{2} (\bf{1}+\bf{S} ) \bf{A}^{(0)}.
\end{equation}

Note that in the elastic limit, $\eta\rightarrow 1$, the amplitudes decouple and Watson's theorem is recovered.  Note also that   (19) is different from the adhoc prescription often used in the literature\cite{NRSX} where the factor $\frac{1}{2} (\bf{1}+\bf{S}) $ is replaced by $\bf{S}^{\frac{1}{2}} $.
\par
Clearly, we can either use (a, b, c) or $(\delta_1, \delta_2, \eta )$ as our fsi parameter set. We have chosen to work with the latter as,  in some sense,  they have a more intuitive meaning. We calculated $A^{(0)}_1 $ and $A^{(0)}_2 $ using $ H_{eff} $ of (3) with the Wilson coefficients given in (5)-(9) in the factorization approximation with Bauer, Stech and Wirbel (BSW)\cite{BSW}
formfactors. We used an effective $q^2=m_b^2/2 $  for illustrative purposes and the following parameters,
\begin{eqnarray}
~& m_u=5MeV,~~m_s=175MeV,~~m_c=1.35GeV,~~m_b=5.0GeV, &~\nonumber \\
~& CKM~ angles : A=0.90, ~~\lambda=0.22,~~ \rho=-0.12,~~\eta=0.34 ~ \cite{AL},& ~ \\
~&  f_D=200 MeV, ~~f_{D_s}=f_{\eta_c}=300 MeV. & ~ \nonumber
\end{eqnarray}

The uncoupled decay amplitudes have a general structure
\begin{eqnarray}
A^{(0)}_1=v_u A^{(1)}_u +v_c A^{(1)}_c \nonumber \\
A^{(0)}_2=v_u A^{(2)}_u +v_c A^{(2)}_c 
\end{eqnarray}
where $v_u=V_{ub}V^*_{us} $ and $v_c=V_{cb} V^*_{cs} $. The tree-level  operators contribute to $A^{(1)}_c $ and $A^{(2)}_c $ in our case.
\par
As we have three different fsi parameters, we have displayed  a sample of our results for the pair of channels, $B^-\rightarrow D^0 D^-_s $ and $ \eta_c K^- $ in 3-dimensional plots in Fig.1.
\par
A recurring feature of our calculations with different elasticities  was that even for  $\eta $ very close to unity , the CP asymmetry induced in $B^-\rightarrow \eta_c K^- $ was of the same order as in $B^-\rightarrow D^0 D^-_s $. This can be understood as follows:  consider the point $\delta_1=\delta_2=0 $ in our plots, and $\eta $ close to its elastic value, $\eta=1-\epsilon $. Then $(1-\eta^2)^{1/2} \approx (2\epsilon )^{1/2} $,  and is quite significant even for $\eta=0.9 $.  As is seen from Eq.(18),
the coupling of the two channels then occurs through a purely imaginary off-diagonal element at the point $\delta_1=\delta_2=0 $. As the CP asymmetry demands that two distinct CKM angles be involved,  it arises from $Tree\otimes Penguin $   couplings between channels 1 and 2 induced via the off-diagonal element $i(2\epsilon )^{1/2} $.
\par
We repeated this calculation for the pair of decays:  $B^-\rightarrow D^0 D^- $  and $\eta_c \pi^- $.   We show  an example  of the results for this pair of channels  in  Fig.2.
\par
To conclude, we have investigated the effect of inelastic fsi on single isospin two-body modes of B decays. We restricted ourselves to modes where in absence of inelastic fsi the CP asymmetry vanishes  and demonstrated that a coupling to a second channel with nonvanishing asymmetry could lead to an asymmetry in the first channel of the same size as in the second.   It is possible that this conclusion would be diluted if more than two channels were coupled as there would be a fortuitious cancellation of the absorptive parts.
\par
This work was partially supported by an individual  research grant to A.N.K from the Natural Sciences and Engineering Research Council of Canada.

\newpage
\begin{flushleft}
{\Large\bf Figure Captions} \\
Fig.1:  (a) CP asymmetry ($\% $) in $B^-\rightarrow \eta_c K^- $ for $\eta=0.9 $  as a function of $\delta_1 $ and $\delta_2 $.\\
~~~~~~(b) CP asymmetry  ($\% $) in $B^-\rightarrow  D^0D^-_s$for $\eta=0.9 $  as a function of $\delta_1 $ and $\delta_2 $.\\
Fig.2: (a) CP asymmetry  ($\% $) in $B^-\rightarrow \eta_c \pi^- $for $\eta=0.9 $  as a function of $\delta_1 $ and $\delta_2 $.\\
~~~~~~(b) CP asymmetry  ($\% $) in $B^-\rightarrow  D^0D^-$for $\eta=0.9 $  as a function of $\delta_1 $ and $\delta_2 $.\\

\end{flushleft}


\begin{thebibliography}{99}
\bibitem{comment}
Though it is often assumed that these two types of strong interaction phases are one and the same, we emphasize that they are distinct.
\end{thebibliography}
\end{document}